\documentclass[10pt, a4paper, twocolumn]{article} % 10pt font size (11 and 12 also possible), A4 paper (letterpaper for US letter) and two column layout (remove for one column)
\usepackage{flushend}

\usepackage{microtype} % Better typography

\usepackage{amsmath,amsfonts,amsthm} % Math packages for equations

\usepackage[svgnames]{xcolor} % Enabling colors by their 'svgnames'

\usepackage[hang, small, labelfont=bf, up, textfont=it]{caption} % Custom captions under/above tables and figures

\usepackage{booktabs} % Horizontal rules in tables

\usepackage{lastpage} % Used to determine the number of pages in the document (for "Page X of Total")

\usepackage{graphicx} % Required for adding images

\usepackage{enumitem} % Required for customising lists
\setlist{noitemsep} % Remove spacing between bullet/numbered list elements

\usepackage[export]{adjustbox}

\usepackage{sectsty} % Enables custom section titles
\allsectionsfont{\usefont{OT1}{phv}{b}{n}} % Change the font of all section commands (Helvetica)

\usepackage{geometry} % Required for adjusting page dimensions

\def\farcs{\hbox{$.\!\!^{\prime\prime}$}}

\usepackage[T1]{fontenc} % Output font encoding for international characters
\usepackage[utf8]{inputenc} % Required for inputting international characters

\usepackage{mathptmx} % Use the XCharter font

\usepackage{fancyhdr} % Needed to define custom headers/footers
\fancypagestyle{firstpage}{ % Page style for the first page with the title
	\fancyhf{}
}

\newcommand{\authorstyle}[1]{{\large\usefont{OT1}{phv}{b}{n}\color{Black}#1}} % Authors style (Helvetica)

\newcommand{\institution}[1]{{\footnotesize\usefont{OT1}{phv}{m}{sl}\color{Black}#1}} % Institutions style (Helvetica)

\usepackage{titling} % Allows custom title configuration

\newcommand{\HorRule}{\color{Black}\rule{\linewidth}{1pt}} % Defines the black horizontal rule around the title

\pretitle{
	\vspace{-30pt} % Move the entire title section up
	\HorRule\vspace{10pt} % Horizontal rule before the title
	\fontsize{32}{36}\usefont{OT1}{phv}{b}{n}\selectfont % Helvetica
	\color{Black} % Text colour for the title and author(s)
}

\posttitle{\par\vskip 15pt} % Whitespace under the title

\preauthor{} % Anything that will appear before \author is printed

\postauthor{ % Anything that will appear after \author is printed
	\vspace{10pt} % Space before the rule
	\par\HorRule % Horizontal rule after the title
	\vspace{20pt} % Space after the title section
}

\usepackage{lettrine} % Package to accentuate the first letter of the text (lettrine)
\usepackage{fix-cm}	% Fixes the height of the lettrine

\newcommand{\initial}[1]{ % Defines the command and style for the lettrine
	\lettrine[lines=3,findent=4pt,nindent=0pt]{% Lettrine takes up 3 lines, the text to the right of it is indented 4pt and further indenting of lines 2+ is stopped
		\color{Black}% Lettrine colour
		{#1}% The letter
	}{}%
}

\usepackage{xstring} % Required for string manipulation

\newcommand{\lettrineabstract}[1]{
	\StrLeft{#1}{1}[\firstletter] % Capture the first letter of the abstract for the lettrine
	\initial{\firstletter}\textbf{\StrGobbleLeft{#1}{1}} % Print the abstract with the first letter as a lettrine and the rest in bold
}

\usepackage[backend=bibtex,style=authoryear,natbib=true]{biblatex} % Use the bibtex backend with the authoryear citation style (which resembles APA)

\usepackage[autostyle=true]{csquotes} % Required to generate language-dependent quotes in the bibliography

\title{Robust modeling of quadruply lensed quasars (and random quartets) using Witt's hyperbola} % The article title

\author{
	\authorstyle{Raymond A. Wynne\textsuperscript{1} and Paul L. Schechter\textsuperscript{1,2}} % Authors
	\newline\newline % Space before institutions
	\textsuperscript{1}\institution{MIT Department of Physics, Cambridge, MA 02139}\\ % Institution 1
	\textsuperscript{2}\institution{MIT Kavli Institute for Astrophysics and Space Research, Cambridge, MA 02139}\\ % Institution 2
	}

\date{\today} % Add a date here if you would like one to appear underneath the title block, use \today for the current date, leave empty for no date

\begin{document}

\maketitle % Print the title

\thispagestyle{firstpage} % Apply the page style for the first page (no headers and footers)

%----------------------------------------------------------------------------------------
%	ABSTRACT
%----------------------------------------------------------------------------------------

\lettrineabstract{We develop a robust method to model quadruply lensed
  quasars, relying heavily on the work of (\cite{witt}), which showed
  that for elliptical potentials, the four image positions, the
  source, and the lensing galaxy lie on a right hyperbola. For the
  singular isothermal elliptical potential, there exists a
  complementary ellipse centered on the source which also maps through
  the four images, with the same axis ratio as the potential but
  perpendicular to it. We first solve for Witt's hyperbola, reducing
  the allowable space of models to three dimensions. We then obtain
  the best fitting complementary ellipse. The simplest models of
  quadruple lenses require seven parameters to reproduce the observed
  image configurations, while the four positions give eight
  constraints. This leaves us one degree of freedom to use as a figure
  of merit. We applied our model to 29 known lenses, and include their
  figures of merit. We then modeled 100 random quartets. A selection
  criterion that sacrifices 20\% of the known lenses can exclude 98\%
  of the random quartets.}

%----------------------------------------------------------------------------------------
%	ARTICLE CONTENTS
%----------------------------------------------------------------------------------------

\section{Introduction}

Astronomers with experience studying quadruple lenses can reliably determine, by examination of relative positions and fluxes, whether a quartet of point sources is lensed. They have trained a neural network, located between their ears, to identify such systems. With the advent of the Gaia probe, quadruple lenses are being discovered at an astonishing rate (\cite{lemon}), and as evidenced by (\cite{delchambre}), there is a clear necessity for robust methods to model these systems. 

It is widely thought that the gravitational equipotentials that produce most quadruply lensed quasars can be reasonably approximated by concentric ellipses (\cite{ellipticalpotentials}). The simplest models for isothermal elliptical potentials include seven parameters, and a quartet of image positions gives eight constraints (\cite{sevenparam}). While that leaves one degree of freedom for use as a figure of merit, it can be difficult to find the best fitting model in that seven dimensional space.

In what follows, we lean heavily on the work of (\cite{witt}), who finds
that for elliptical potentials, the positions for all four images, the center
of the lens, and the projected (but unobservable) position of the
quasar all lie on an hyperbola whose asymptotes align with the
potential's major and minor axes.  The hyperbola gives us
the position angle and restricts the positions of the lens and the
source to a one dimensional locus.  It reduces the dimensionality of
the space to be searched from 7 to 3.

We also show that for the specific case of the singular isothermal elliptical potential, there exists an ellipse mapping through all 4 image positions, whose minor axis is aligned with the major axis of the potential, has an axis ratio inverse to the axis ratio of the potential, and is centered on the source.

\begin{figure*}[t!]
    \centering
    \includegraphics[width=16cm]{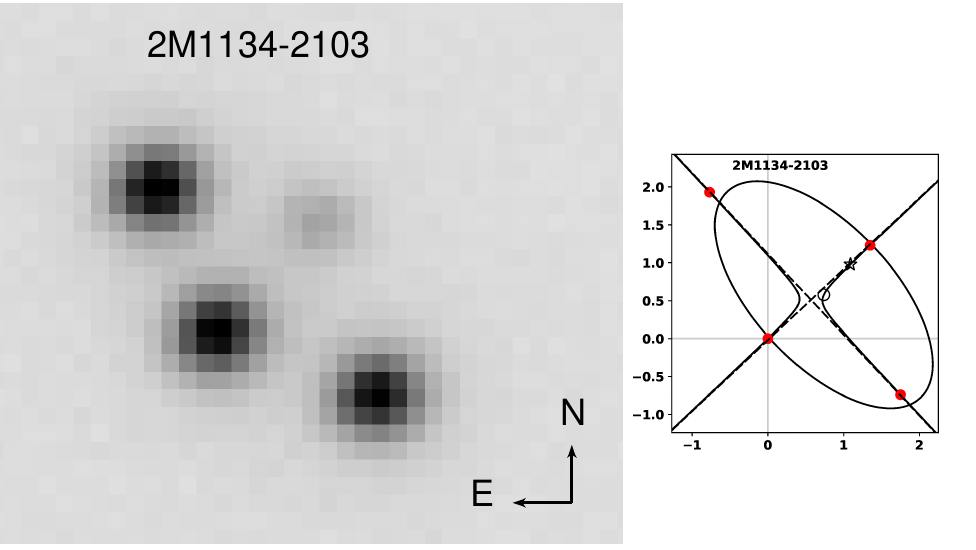}
    \caption{Quadruple lens system 2M1134-2103, with both images at the same scale. On the left is a Sloan i filter image taken from the ATLAS survey, and the scale of this figure is 0\farcs21 per pixel. On the right is the plot our model produces for 2M1134-2103. The red dots are the image positions, the circle is the source position, and the star is the galaxy position. The parameter values are $b = 0\farcs92, q = 0.491, \theta = 133.1^{\circ}$ with $\sigma_{\ln{b}} = 0.071.$}
    \label{fig:2M1134}
\end{figure*}

We then search along Witt’s hyperbola for the source position and axis ratio that minimize the scatter in the lensing strengths determined by the positions of the four quasar images. We use this scatter as our figure of merit.

In Section 2, we briefly explain gravitational lensing, and discuss the importance of quadruple lenses. In Section 3, we give background into the geometry associated with the quadruple systems, and explain our method of solving for the source position and axis ratio by minimizing the scatter in the lens strength. In Section 4, we compute our proposed figure of merit with a sample of spectroscopically confirmed quadruply lensed quasars. In Section 5, we analyze random quartets, and compare them to the known figures of merit. In Section 6, we discuss the results of the previous sections, and how they might be used to accept or reject the lensing hypothesis.

%-------------------------------------------------------------------------------------

\section{Background}

\subsection{Gravitational Lensing}
Albert Einstein's theory of General Relativity  represents gravity as the warping of space-time. Light propagates along paths called geodesics, which are a generalization
of the notion of straight lines to curved spaces.  When gravitational
fields are not strong, one can treat the propagation of light
as if there were an index of refraction proportional
to the gravitational potential (\cite{refraction}).

From Fermat's principle, it is known that light follows a path that is stationary in time. For everyday situations, this is the single path that takes the least amount of time for light to travel (\cite{BlandfordNarayan}). However, on the astronomical scale, a massive body such as a galaxy can warp space to the degree that multiple paths satisfy this stationary criteria, and multiple images of one source object appear. In this way, the galaxy acts as a lens," although the French have a better name for the phenomenon: \textit{mirage gravitationnel} (\cite{mirage}).

\begin{figure*}[t!]
    \centering
    \includegraphics[width=16cm]{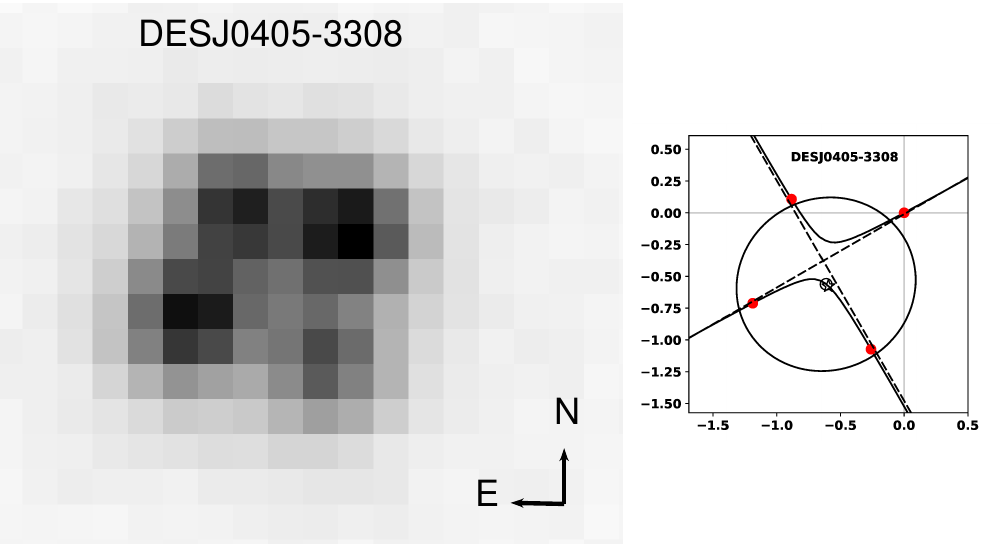}
    \caption{Quadruple lens system DESJ0405-3308, with both images at the same scale. On the left is a Sloan i filter image taken from the DES survey, and the scale of this figure is 0\farcs26 per pixel. On the right is the plot our model produces for DESJ0405-3308. The red dots are the image positions, the circle is the source position, and the star is the galaxy position. The parameter values are $b = 0\farcs67, q = 0.944, \theta = -120.6^{\circ}$ with $\sigma_{\ln{b}} = 0.130.$}
    \label{fig:DESJ0405}
\end{figure*}

In gravitational lensing, there are three important phenomena and their corresponding equations: time delay, deflection, and distortion. The first deals with the difference in travel time between different paths, the second with the degree to which light gets bent towards the observer, and the final with how the images are magnified and distorted. In this paper, we will only be dealing with deflection, and the associated lens equation,
\begin{equation}
    (\vec{u} - \vec{u_s}) = \vec\nabla{\psi},
\end{equation}
where $\vec{u_s}$ is the position at which the source would be in the absence of lensing, and $\psi$ is the projected, two-dimensional gravitational potential obtained by integrating the three-dimensional potential along the line of sight. The solutions of the lens equation, $\vec{u_i}$, give the positions of the images of the source. The $\vec{u} - \vec{u_s}$ term is also known as the deflection, since it is the difference between the observed position and the position the source would have had sans lensing. Therefore, the stronger the gradient in the potential, the greater the deflection (\cite{BlandfordNarayan}).

% \begin{figure*}[t!]
%     \centering
%     \includegraphics[width=16cm]{DESJ0405.pdf}
%     \caption{Quadruple lens system DESJ0405-3308, with both images at the same scale. On the left is a Sloan i filter image taken from the DES survey, and the scale of this figure is 0\farcs26 per pixel. On the right is the plot our model produces for DESJ0405-3308. The red dots are the image positions, the circle is the source position, and the star is the galaxy position. The parameter values are $b = 0\farcs67, q = 0.944, \theta = -120.6^{\circ}$ with $\sigma_{\ln{b}} = 0.130.$}
%     \label{fig:my_label}
% \end{figure*}

\subsection{Quadruple Lenses}

Configurations with four images, known as quadruple lenses, are of particular interest. For example, they permit measurements of time delays (\cite{timedelays}), put constraints on cosmological parameters such as the Hubble constant (\cite{cosmoparam}), allow study of the structure of quasars (\cite{quasarstructure}), offer research on the stellar content of the lensing galaxy through microlensing (\cite{stellarcontent}), and can be used as a probe for gas clouds along the line of sight (\cite{gasclouds}).
\raggedbottom

Nearly four dozen quadruply lensed quasars have been discovered since PG1115+080 (\cite{pg1115}), many in recent years. Gaia, whose mission is to create a detailed three-dimensional map of the Milky Way, offers a way to detect these systems (\cite{GAIAmission}); it is expected to discover approximately 2900 quasars, with around 80 having two or more images (\cite{finetsurdej}).
\raggedbottom

All of the many applications of quadruply lensed quasars require a model for the gravitational potential. In the gravitational lensing literature, a frequently used model is the elliptical potential. Witt argues that equation (4) applies equally well
to  a singular isothermal sphere with external shear  $\gamma$,
substituting  $1+\gamma/1-\gamma$ for $1/q^2$ in equation (4).
This is true for the special case in which the shear term in
the potential is centered on the source (\cite{shear}).  When the shear term is centered on the galaxy the hyperbola is offset.  One way or the other,
one might model a singular isothermal sphere with shear following the approach developed below for the singular isothermal elliptical potential.

%--------------------------------------------------------------------------------------

\section{Method}
\subsection{Witt's Hyperbola}

In Witt's paper, he assumes an elliptical potential of the form $\psi = f(r)$, where $r = x^2 + y^2/q^2$ is the equation of an ellipse with semi-major axis aligned with the x-axis of the coordinate system, and where $f$ is an arbitrary function that describes the variation in spacing of the elliptical equipotentials. For the purposes of this paper, we will use the singular isothermal elliptical potential
\begin{equation}
\begin{split}
    \psi = b\sqrt{(x - x_g)^{2} + (y - y_g)^{2}/q^2}
    = bt
\end{split}
\end{equation}
where $b$ is the lens strength, $(x_g, y_g)$ is the position of the lensing galaxy, $q (<1)$ is the axis ratio of the potential, and $t = \sqrt{r}$. Note that $t$ plays the role of a distance from the galaxy. The singular isothermal elliptical potential, hereafter SIEP, is a good approximation to the actual potentials produced by the lensing galaxy (\cite{ellipticalpotentials}).
 
The lens equation for this potential is the following,
\begin{multline}
    \vec{u} - \vec{u_s}= \cfrac{1}{t}\cfrac{d\psi}{dt}\Big[(x - x_g)\hat{x} + \frac{(y - y_g)}{q^2}\hat{y}\Big],
\end{multline}
where $(x_s, y_s)$ is the (unobservable) position of the source.
Witt uses only the direction of this equation, taking the ratio of the y component of the displacement vector to the x component of the
displacement vector to get

\begin{equation}
    \cfrac{(y - y_s)}{(x - x_s)} = q^{-2}\cfrac{(y - y_g)}{(x - x_g)}
\end{equation}

\begin{figure*}[t!]
    \centering
    \includegraphics[width=16cm]{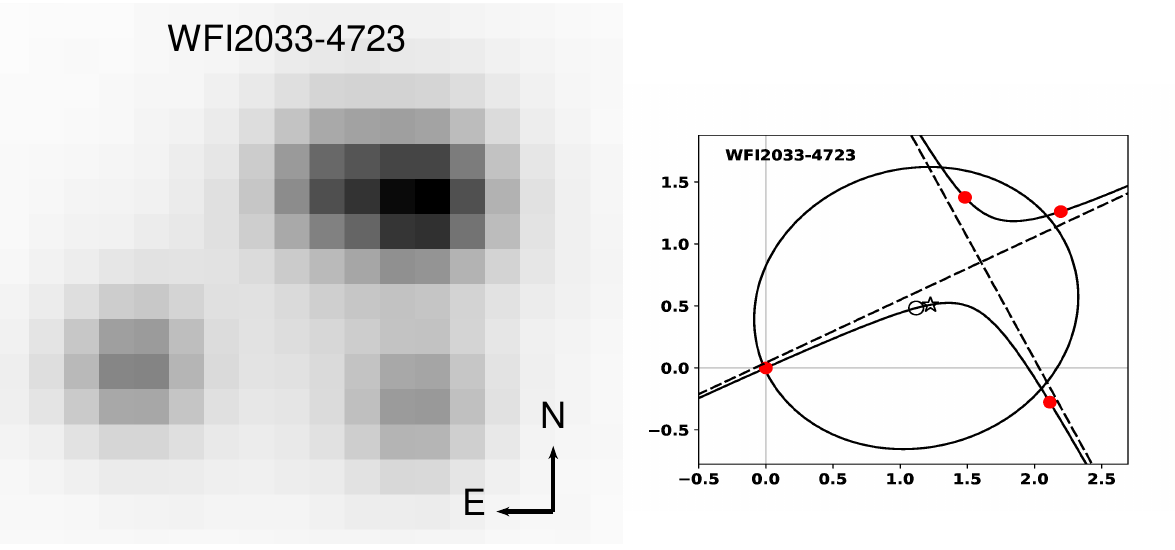}
    \caption{Quadruple lens system WFI2033-4723, with both images at the same scale. On the left is a Sloan i filter image taken from the DES survey, and the scale of this figure is 0\farcs26  per pixel. On the right is the plot our model produces for WFI2033-4723. The red dots are the image positions, the circle is the source position, and the star is the galaxy position. The parameter values are $b = 1\farcs12, q = 0.906, \theta = -63.1^{\circ}$ with $\sigma_{\ln{b}} = 0.118.$}
    \label{fig:WFI2033}
\end{figure*}

Cross multiplication of (4) gives an equation of the form 
\begin{equation}
    xy + Ax + By = C,
\end{equation}
(the coefficients being determined by the galaxy and source positions and the axis ratio), which happens to be a right hyperbola with asymptotes aligned with the x and y axes. It is clear that, along with the four images, the galaxy and source must lie on this curve, since they satisfy the equation above.

We can with no loss of generality take our coordinate system to
be one in which the asymptotes of the hyperbola are not merely
aligned with the axes but coincident with them, in which case:
\begin{equation}
    xy = W^{2},
\end{equation}
where W is a constant with units of length.

One must apply an arbitrary rotation and translation to Witt's hyperbola to give it the form that it takes in the observed coordinate system. If one shifts the images in the observed system system such that one image is at the origin, the equation for the hyperbola takes a simpler form, and is as follows:
\begin{equation}
    c_1x'^{2}+c_2x'y'-c_1y'^{2}+c_3x'+c_4y' = 0.
\end{equation}

We observe only the four images.  The position of the source is unknown, the position of the lensing galaxy is often unknown, and even when the position of the lensing galaxy is known, the equipotentials will not have the same shape since the galaxy may not have the same orientation. The image positions are measured in a coordinate system that is rotated and translated by unknown amounts from the coordinate system in which Witt's hyperbola takes its simplest form.

Gravitational lensing requires seven parameters to model a system: the orientation $\theta$, the two galaxy coordinates $(x_g, y_g)$, the two source coordinates $(x_s, y_s)$, the lens strength $b$, and the axis ratio $q$.  Witt's hyperbola has immense power in modeling systems. It not only gives the orientation, but also describes the position of the galaxy and source by one coordinate, not two. Therefore, Witt's hyperbola reduces the dimensionality of the space to be searched from 7 to 3.

Four observed image positions suffice to determine the four unknown
coefficients, $c_1 - c_4$, that uniquely describe the hyperbola.  If the
lensing potential is {\it not} elliptical the four images will still
give a hyperbola, but we would not expect the source or the galaxy to
lie on it.  Indeed any four points will produce a right hyperbola, so
the hyperbola, by itself, does not tell us if four images are in fact
lensed or whether the potential producing them is elliptical.

\subsection{A Complementary Ellipse}

So far, we have only used the direction of the lens equation. Taking the magnitude squared of each side of (3) and a little algebra gives the equation
\begin{equation}
    (x - x_s)^2 + q^2(y - y_s)^2 = b^2
\end{equation}
This equation is important in that for the SIEP, the images appear at the intersection of Witt's hyperbola and an ellipse centered on the unlensed source position, with minor axis aligned with the major axis of the potential.

The four image positions are enough to determine the four parameter values of equation (8). But as
in the case of the hyperbola, {\it any} four points will produce
an ellipse once we specify the orientation of the axes. However, unless the images are precisely those of a singular isothermal
elliptical potential, the hyperbola determined by equation (7) and the ellipse determined by equation (8) are not
consistent with each other.  The source position, as determined
from the ellipse, will not, in general, lie on the hyperbola.

\subsection{Modeling the Singular Isothermal Elliptical Potential}

As we wish to construct a self-consistent model, we chose one that is consistent with the hyperbola determined from the four images.  We then find an ellipse of the form of equation (8), with the source on
our hyperbola, that comes close to passing through the four images.

It is evident from equation (8) that for variable sourceH position and axis ratio, each image will have an associated lens strength $b_i$. For perfect source position and axis ratio, the $b_i$ will be equal, since the configuration is the result of one lensing strength. Therefore, minimizing the standard deviation of the logarithm of the lens strength,
\begin{equation}
        \sigma_{\ln{b(x_s, q)}} = \sqrt{\Big\langle{(\ln{b_i} - \big\langle{\ln{b}}\big\rangle)^{2}\Big\rangle}},
\end{equation}
where $\big\langle{\ln{b}}\big\rangle = \frac{1}{4}\sum_{i=1}^{4}\ln{b_i}$, will give the source position and axis ratio that is in this sense the
best fit. In the coincident coordinate system, this is a simple, 2-dimensional minimization problem, since the y-coordinate is given by $\frac{W^{2}}{x}$. As we show in Appendix A, once the source position and axis ratio are known, the galaxy position is immediately determined. In Appendix B, we show how to determine which branch of the hyperbola the source and galaxy lie on.

Considering the singular isothermal elliptical potential is only a model, albeit a very useful one, the minimum of the scatter will not be at zero. However, this is good; we can use this minimum value as a figure of merit. It can be used to determine how closely a system resembles the singular isothermal elliptical potential, offering insight as to whether a system is a lens or not.

%---------------------------------------------------------------------------------------

\section{Known Lenses}

To test our method, we examined 29 known quadruple lenses and recorded their figures of merit. The average value for the figure of merit was $\langle\sigma_{\ln{b}}\rangle = 0.0531$, and the standard deviation was $\sigma_{\sigma_{\ln{b}}} = 0.0468$. 

We also include three spectroscopically determined quadruply-lensed quasars, and their plots determined by our program. As seen in the plot, the hyperbola goes through all four image positions, along with the source and galaxy positions. In minimizing the scatter, you determine a source position and axis ratio, which in turn determines the ellipse. For the specific case of the singular isothermal elliptical potential, this ellipse will map through the four points; however, since the potential is only a model, this ellipse will not go perfectly through all four images for real systems, but the deviation from the image positions and the ellipse gives a visual for the scatter.

\begin{table}[!h]
\centering
\begin{tabular}{|p{1.05in}|p{.85in}|p{.7in}|}
 \hline\hline
 System & Figure of Merit: $\sigma_{\ln{b}}$ & Axis Ratio\\
 \hline\hline
 
 GRAL1131-4419&	            0.002&		0.901\\
 SDSSJ1138+0314&		    0.004&		0.818\\
 ATLAS0259-1635&		    0.009&		0.892\\
 HE0435-1223&		        0.010&		0.861\\
 HE1113-06412&		        0.015&		0.924\\
 PSJ0147+4630&		        0.015&		0.711\\
 RXJ1131-1231&		        0.016&		0.746\\
 HS0810+2554&		        0.018&		0.896\\
 WGD2100-4452&		        0.019&		0.853\\
 Q2237+030&		            0.020&		0.876\\
 SDSSJ1251+2935&	        0.027&		0.769\\
 WGD2038-4008&		        0.033&		0.771\\
 WFI2026-4536&	        	0.036&		0.765\\
 MG0414+0534*&		        0.038&		0.658\\
 DESJ0408-5354*&		    0.041&		0.818\\
 DESJ0924+0219&		        0.044&		0.910\\
 RXJ0911+0551*&		        0.045&		0.563\\
 B0712+472&		            0.049&		0.883\\
 2M1310-1714&		        0.051&		0.953\\
 WG0214-2105&		        0.052&		0.782\\
 PG1115+080&		        0.053&		0.769\\
 WISE2344-3056&		        0.070&		0.862\\
 2M1134-2103&	        	0.071&		0.491\\
 PS0630-1201*&	        	0.082&		0.401\\
 SDSSJ1330+1810&	    	0.091&		0.991\\
 WFI2033-4723*&	        	0.118&		0.906\\
 H1413+117&		            0.123&		0.801\\
 DESJ0405-385&	        	0.130&		0.944\\
 HE0230-2130*&	        	0.220&		0.707\\
 
 \hline 
\end{tabular}
\caption{List of 29 known quadruples, with their associated figures of merit and axis ratios. * denotes a two-lens system.}
\label{table:1}
\end{table}

Two systems above caused trouble for our program: HE0230-2130 and PS0630-1201; the former having the largest scatter, and the latter having a galaxy position that lies outside the ellipse. Both of these systems have two lensing galaxies, so their potentials deviate from the elliptical model more so than systems with one lensing galaxy, which might explain their issues.

There is a natural physical limit q > 0.5 to the singular isothermal elliptical potential as an edge-on infinitely flat galaxy with a flat rotation curve produces equipotentials with q = 0.5 (\cite{axisratiolimit}).  Thus the models for systems with $q \le 0.5$ are very suspect. The axis ratio q = 0.49 derived for 2M1134-2103
argues against our model.  (\cite{2m1134}) find that the system is
better modeled as a singular isothermal sphere with external shear
$\gamma=0.34$, among the highest known for quadruply lensed quasars.

As Witt mentioned, there is ambiguity in determining the orientation of the system. The correct orientation has the asymptotes of the hyperbola aligned with the axes, but there are two unique orientations with that property. In Appendix C, we sort out this issue.

%----------------------------------------------------------------------------------------

\section{Random Quartets}

\begin{figure*}[t!]
    \centering
    \includegraphics[width=15cm]{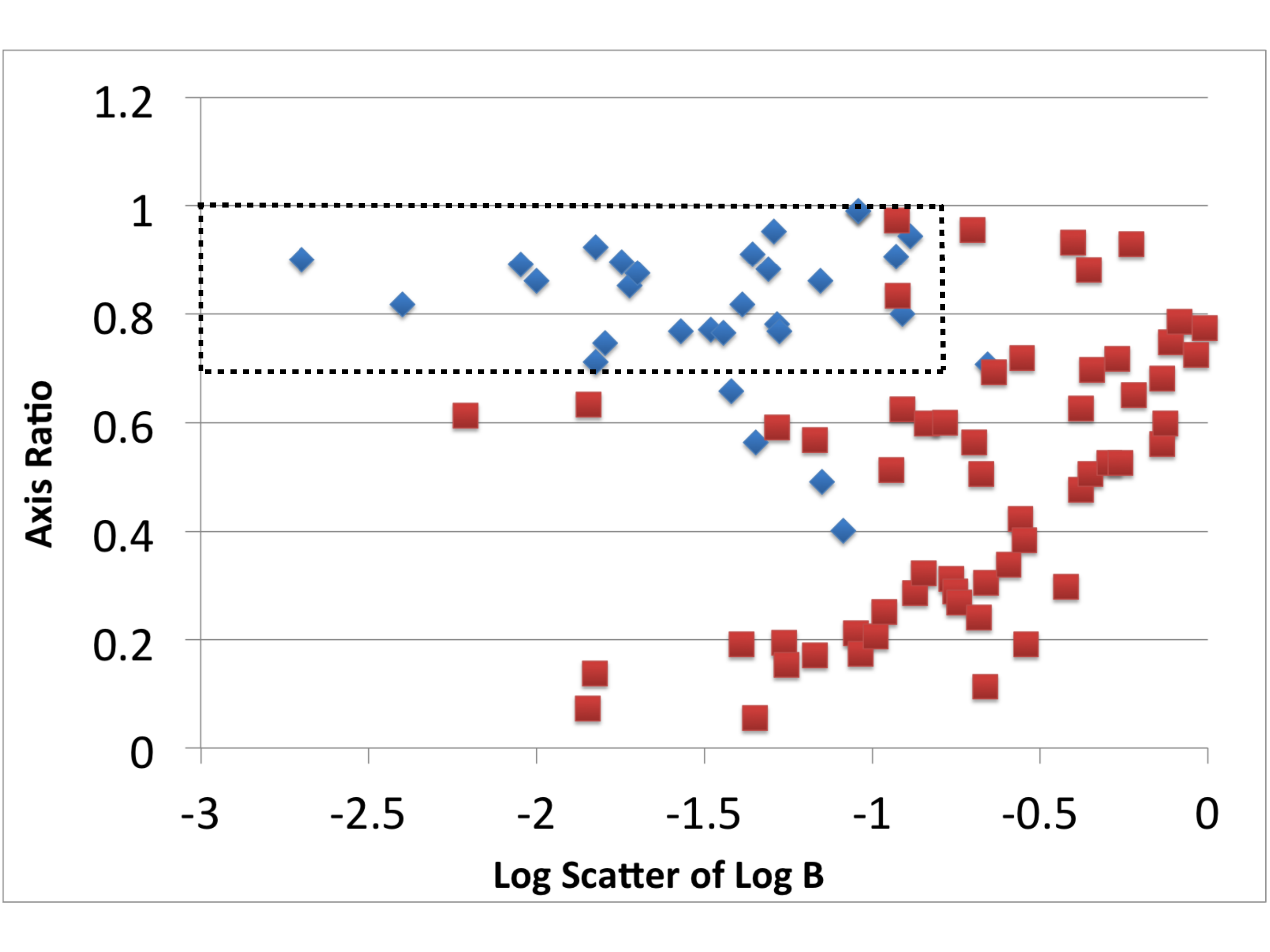}
    \caption{Scatter plot of the logarithm of the figure of merit vs the axis ratio. Twenty-nine known lenses are in blue, and fifty-six random quartets are in red. The dashed box catches 24 of the 29 known lenses, as well as 2 of the random quartets.}
    \label{fig:scatter}
\end{figure*}

To see how well our figure of merit discriminated between genuine lenses and random quartets, we generated 100 random quartets within the unit circle and applied our program to them. We rejected the cases that had three or more images on one branch of the hyperbola.

In Figure 4, we plot the logarithm of the scatter versus the axis ratio for our known system and random quartets that passed our criteria above. We hoped for more separation between the real and random quartets, but every model has it's downsides. The dashed box catches 24 of the 29 known lenses, as well as 2 of the 100 random quartets. So roughly, our model has a 2\% false positive rate. Two percent of 80,000, which is roughly the number of quartets identified by (\cite{delchambre}), is 1,600; in other words, a lot of eyeballing is still required.

The systems that lie below the box include RXJ0911+0551 and 2M1134-2103 for which external shear rather than the ellipticity of the lens is thought to dominate the quadrupole term in the potential.  MG0414+0534 and PS0630-1201, also below the box, are systems for which a second lensing galaxy contributes significantly to the potential. HE0230-2130, another double lens, lies to the right of the box. While it's a shame to lose these systems, they are all pretty different from our adopted model.

It is clear from the comparison of our lensed systems and our random
quartets that positions alone do not permit perfect discrimination
between the two.  If this discrimination does not suffice for
some particular task, there is additional information that might be
brought to bear on improving it -- the relative
fluxes of the four images.

In the absence of micro-lensing, the relative fluxes that our model
parameters predict for the four images can be obtained from the
distortion equation.  These cannot be used directly, because
micro-lensing is expected to be universal in quadruply lensed quasars
(\cite{microlensing}).  (\cite{yaha}) have shown how one might
take micro-lensing into account in assessing the likelihood that
observed flux ratios are consistent with the lens model and
microlensing.  One would then still need to decide how to combine the
present astrometric discrimination and the photometric figure of merit.

%----------------------------------------------------------------------------------------

\section{Conclusion}

We discussed the basics of gravitational lensing, and the phenomenon of quadruple lenses. We then developed a method for modeling quadruply lensed quasars through the use of Witt's hyperbola and the complementary ellipse, as well as assigning a figure of merit to potential systems. We applied this method to 29 known quadruply lensed systems and 100 random quartets, and included their figures of merit. For future systems, this figure of merit can help astronomers determine if a newly discovered system is the product of gravitational lensing, or merely a random configuration.

{\it Acknowledgements:} We thank Professor Alar Toomre for asking how we could tell quadruply lensed quasars from random quartets. We thank Professor Chuck Keeton of Rutgers for
setting us straight about ellipses early in this effort and his
pointing us toward Witt's work.  Finally, we thank the MIT Undergraduate
Research Opportunities Program for support.

%	BIBLIOGRAPHY
%----------------------------------------------------------------------------------------

\printbibliography[title={References}] % Print the bibliography, section title in curly brackets

%----------------------------------------------------------------------------------------

\makeatletter
\def\@seccntformat#1{%
  \expandafter\ifx\csname c@#1\endcsname\c@section\else
  \csname the#1\endcsname\quad
  \fi}
\makeatother

\section{Appendix A}

In this appendix, we show how to determine the galaxy position from the source position and axis ratio. Using the direction of the lens equation, Witt derived equation (4). This equation gives us a right hyperbola mapping through the four images, the source, and the galaxy positions. Essentially, this hyperbola gives us the possible locations for image positions given the configuration of the system. For simplification, moving the y-terms to the left and x-terms to the right, one recovers

\begin{equation}
    \frac{(y - y_s)}{(y - y_g)} = q^{-2}\frac{(x - x_s)}{(x - x_g)}
\end{equation}

In the coincident coordinate system, the y-coordinate is given by $W^2/x$, so as x goes to zero, the left hand side approaches unity while the right hand side approaches $q^{-2}\cfrac{x_s}{x_g}$. Therefore, for perfect source position and axis ratio, the x-coordinate of the galaxy position is given by

\begin{equation}
    x_g = \frac{x_s}{q^2},
\end{equation}
and since the galaxy also lies on the hyperbola, the y-coordinate is immediately determined.

%-------------------------------------------------------------------------------------------

\section{Appendix B}

In the coincident coordinate system, the branches of the hyperbola are separated by the x and y axes. The question arises as to which branch the galaxy and source will lie on. To determine this, we minimize the scatter, subject to the constraint that the source lies within the image configuration, on both branches of the hyperbola. The branch that the source which minimizes the scatter lies on is determined to be the correct branch.

%--------------------------------------------------------------------------------------------

\section{Appendix C}

The hyperbola gives us two unique orientations for the system. However, the ellipse determines which of these systems is correct. Looking at equation (8), it is easy to see that coefficient for $y^2$ is simply $q^2$. Thus, determining the equation for the ellipse using the four image positions determines the axis ratio. Therefore, if q>1, you know you are in the wrong orientation. If you happen to be in the orientation where q>1, rotating by $\frac{\pi}{2}$ gets you into the correct orientation. 

However, since the ellipse determined by the image positions is only accurate for near-perfect systems, it might not give the orientation which minimizes scatter. Therefore, we use our method for four possible orientations: observed, rotated by +/- $\frac{\pi}{2}$, and rotated by $\pi$. The orientation which gives the smallest scatter is determined to be the correct orientation. We then rotate the found source and galaxy positions back into the observed frame, if need be.

\end{document}